\documentstyle[preprint,aps]{revtex}
 \begin{document}
 \draft
 \title{Comparison of Bond Character in Hydrocarbons and Fullerenes}

 \author{D.W. Snoke}

 \address{Department of Physics and Astronomy, University of Pittsburgh,
 3941 O'Hara St., Pittsburgh, PA 15260, USA}

 \author{M. Cardona}

 \address{Max-Planck Institut f\"ur Festk\"orperforschung,
 Heisenbergstr. 1, 70506 Stuttgart, Germany}

 \author{S. Sanguinetti and G. Benedek}

 \address{INFM -- Dipartimento de Fisica dell'Universit\'a, Via Celoria 16,
 20133 Milano, Italy}

 \maketitle

 \begin{abstract}
 We present a comparison of the bond polarizabilities for carbon-carbon
bonds in hydrocarbons and fullerenes, using two different models for the
fullerene Raman spectrum and the results of Raman measurements on ethane
and ethylene. We find that the polarizabilities for single bonds in
fullerenes and hydrocarbons compare well, while the double bonds in
fullerenes have greater polarizability than in ethylene.
 \end{abstract}

 \newpage
 Countless experiments have used Raman spectroscopy to identify the
vibrational spectrum of molecules.  This use of Raman scattering inherently
assumes (1) that the process (photon in $\rightarrow$ photon  out + phonon)
is symmetry allowed, and (2) that the {\em change of the electric
 susceptibility} of the molecule, due to the particular molecular
distortion associated with the phonon, is large enough to give an
effect.\cite{LSS1} If the polarizability of the molecule does not change
substantially during a particular vibration, then even if the scattering
process is symmetry allowed, Raman signal will appear only very weakly for
that phonon. The {\em ab initio} estimation of these polarizability changes
 is difficult, however.

 The ``bond polarizability'' model provides a greatly simplifying assumption.
 This model sets the polarizability of a molecule (or molecular solid)
 equal to the sum of polarizabilities of the individual bonds. Bonds are
 treated as independent clouds of electrons between the atoms, and the
further simplifying assumption is made that the polarizability of these
bonds depends only on their length and their total charge. Making this
assumption, not only the Raman line {\em frequencies} can be fit to theory,
but also the relative {\em intensities}.

 This model has the benefit of providing a quantifiable measure of the basic
 characteristic of bonds.  In what way can we say that a double bond
between two carbon atoms in one molecule is similar to or different from a
double bond of carbon atoms in a different molecule, or in a solid? How
much do particulars of the electron distribution really affect the basic
character of bonds? If we can use Raman intensities to measure the
polarizabilities of the two bonds, and we find them nearly the same, we can
justifiably say they have similar character.

 The carbon-carbon bond, ubiquitous in biological systems, polymers, fuels,
and composite materials, has received tremendous attention. Early work
 \cite{burstein} showed that the bond polarizability model works well for
 carbon; more recent work has applied this model to hydrocarbons
 \cite{montero1,montero2,montero3}, and carbon in graphite and diamond
 solids\cite{bilz}. Following the discovery of carbon fullerenes, there has
 obviously been interest in whether the carbon-carbon bonds in fullerenes
have character similar to carbon-carbon bonds in other molecules.

 Two previous works addressed this question using different approaches. In
the first \cite{snoke1}, a next-nearest-neighbor force constant model with
eight spring constants was used to reproduce the vibrational spectrum of
the C$_{60}$ molecule, similar to the 20-spring-constant model used by
Al-Jishi and Dresselhaus \cite{al-jishi} to model the vibration spectrum of
graphite. The icosahedral symmetry of the C$_{60}$ molecule was explicitly
invoked in order to block diagonalize the force constant matrix, and force
constants similar to those in diamond and graphite were found to give a
good fit to the C$_{60}$ IR and Raman vibrational spectrum. Next, the
phonon eigenvectors found from this fit were used in a five-parameter bond
polarizability model
\cite{snoke1} in order to fit Raman intensities of C$_{60}$ reported early
in the literature \cite{bethune}. At the time of those calculations, all
the available Raman data were obtained with visible lasers, which turned
out to have frequency near to the electronic resonance of the fullerenes
\cite{sinha}. Ref. \cite{snoke1} made a rough correction for the effect of
this resonance, but the bond polarizabilities found from a fit of the Raman
intensities did not compare well with those from hydrocarbons.

 In a different approach \cite{sang}, use has been made of a four-parameter
bond charge model to fit the vibrational spectrum of C$_{60}$ and C$_{70}$.
Having obtained a successful fit to the vibration frequencies, this model
 automatically provides an estimate of the bond polarizabilities without
 additional fitting. This is because a bond charge model makes certain
 assumptions about how the charge of the bonds redistributes under
distortion, which is exactly what gives a change in polarizability.  No
comparison to hydrocarbon polarizabilities was made at that time.

 In this paper, we wish to directly compare these two models to the
hydrocarbon data, using an updated fit of the model of Ref. \cite{snoke1}
to Raman intensity data for C$_{60}$ taken far from resonance with a Nd:YAG
laser \cite{chase}. We find that a consistent  picture arises from this
comparison.

 In the bond polarizability model, the polarizability of each bond is written as
 \begin{equation}
 \tilde{\alpha}  = \left| \begin{array}{lll}
 \alpha_{\bot}+\alpha_{\bot}'\cdot d\ell\\
 & \alpha_{\bot}+\alpha_{\bot}'\cdot d\ell\\
 & & \alpha_{\|}+\alpha_{\|}'\cdot d\ell\\
 \end{array}\right|
 \end{equation}
 where the z axis is along the bond, and $d\ell$ is the change in length of
the bond. This leads to four parameters for each bond, namely the isotropic
part  $2\alpha_{\bot} + \alpha_{\|} \equiv {\cal P}$, and its first
derivative, and the anisotropic part, $\alpha_{\|} - \alpha_{\bot} \equiv
{\cal Q}$, and its
 first derivative. (In theory, $\alpha_{\bot}$ could have different values
 along the $x$ (in-plane) and $y$ (out-of-plane) directions, but the data
do not warrant such a distinction here.) Because the constant isotropic
part just contributes to the overall dielectric constant, this leaves three
parameters that contribute to the Raman intensities. Since the single and
double bonds can have different character, six parameters should actually
be used.  Usually, however, the absolute  intensities of the Raman lines
are not measured, and therefore the absolute values of these parameters
cannot be determined. This leads to five ratios among the parameters for
the relative intensities of the Raman lines. By symmetry, the two $A_g$
Raman lines of
 C$_{60}$ depend only on the two isotropic
 parameters ${\cal P}_s$ and ${\cal P}_d$ for the single and double bonds,
 respectively, and the eight $H_g$-symmetry lines depend only on the other
four parameters.

 The results of the fit of the model of Ref. \cite{snoke1} to the
off-resonant Raman data for C$_{60}$ \cite{chase} to the Raman data are
shown in Table 1.  Table 2 shows the ratios of polarizability parameters
obtained from this fit and from the bond-charge model \cite{sang}. Although
the bond charge model in principle does not need a fit for the
 polarizability parameters, in practice the parameter ${\cal Q}'/{\cal Q}$ is
 not well determined by the fit to the vibration spectrum, because the
 anharmonic part of the interatomic Keating potential is not well known
 \cite{anas}. In order to best fit the $A_g$ line intensities, a value of
 ${\cal Q}'/{\cal Q} = 0.43 \AA = 0.3r_0$  was used in the bond charge model
 of Ref. \cite{sang} (where the interatomic spacing in C$_{60}$ is $r_0 =
1.4 \AA$ \cite{shi}. )

 As seen in this Table, the parameters obtained in these two very different
 ways agree remarkably well. These are compared to the experimental values
from ethane (CH$_6$, with one single carbon-carbon bond) and ethylene
(CH$_4$, with one double carbon-carbon bond.)
 This comparison indicates that in both of these models
 the polarizability of the double bonds in C$_{60}$ is about twice
 that of the double bond in ethylene. The bond charge model suggests that
 this result has a physical basis in the way the charge on the double bonds
 redistributes.

 {\bf Acknowledgements}. We thank J. Men\'endez, S. Montero and G. Onida
 for helpful discussions.
We acknowledge partial financial support from  Italian  Consiglio
Nazionale    delle   Ricerche   through   project:    ``Crescita,
caratterizzazione e propriet\'a di strutture fullereniche".

 \begin{table}
\renewcommand{\baselinestretch}{1}\huge\normalsize
 \begin{tabular}{lll}
 \multicolumn{3}{c}{Table 1. Raman Intensities}\\ \hline
 \\
 Frequency &  Relative &  5-parameter \\
 \cite{bethune,chase} &  Intensity \cite{chase} & Fit Intensity \\
 \\
\multicolumn{3}{l}{A$_{g}$ modes:} \\
 496 & 100 &  100\\
 1470 & 92 &  92\\
 \\
 \multicolumn{3}{l}{H$_{g}$ modes:} \\
 273 & 86 & 86\\
 437 & 13 & 11 \\
 710 & 7 &  3\\
 774 & 20 & 12 \\
 1099 & 11 &  16\\
1250 & 10 &  3\\
 1428 & 6 & 5\\
 1575 & 12 & 14\\
 \\
 \end{tabular}
 \end{table}

 \begin{table}
 \begin{tabular}{llll}
 \multicolumn{4}{c}{Table 2. Bond Polarizability Parameters}\\ \hline
 & hydrocarbon  & Ref. \protect\cite{snoke1} model fit & Bond charge \\
  & data: \protect\cite{montero1,montero2} & to Ref. \protect\cite{chase} data &
 model
 \protect\cite{sang}\\
 ${\cal P}_s'/{\cal Q}_s'$ & 1.35 & 1.31 & 1.56\\
 ${\cal Q}_s/{\cal Q}_s'r_0$ & 0.41 & 0.34 & 0.31\\
 ${\cal Q}_d'/{\cal Q}_s'$ & 1.13 & 2.92 & 2.96\\
 ${\cal P}_d'/{\cal Q}_s'$ & 2.81 & 4.97 & 4.68\\
 ${\cal Q}_d/{\cal Q}_s'r_0$ & 0.46 & 0.98 & 0.93\\
 \end{tabular}
 \end{table}

 \end{document}